# Atomic-scale imaging of graphene nanoribbons on graphene after polymer-free substrate transfer

Amogh Kinikar[a], Feifei Xiang[a], Lucia Palomino Ruiz[a,b], Li-Syuan Lu[c], Chengye Dong[d], Yanwei Gu[e#], Rimah Darawish[a,f], Eve Ammerman[a], Oliver Gröning[a], Klaus Müllen[e,g], Roman Fasel[a,f], Joshua A. Robinson[c,d,h], Pascal Ruffieux[a], Bruno Schuler[a], Gabriela Borin Barin[a*]

[a]nanotech@surfaces laboratory, Empa - Swiss Federal Laboratories for Materials Science and Technology, 8600 Dübendorf, Switzerland.

[b]Departamento de Química Orgánica, Facultad de Ciencias, Unidad de Excelencia de Química Aplicada a Biomedicina y Medioambiente (UEQ), Universidad de Granada, 18071 Granada, Spain

[c]Department of Materials Science and Engineering, The Pennsylvania State University, University Park, Pennsylvania 16082, USA

[d]Two-Dimensional Crystal Consortium, The Pennsylvania State University, University Park, Pennsylvania 16802, USA

[e]Max Planck Institute for Polymer Research, 55128 Mainz, Germany

[f]Department of Chemistry, Biochemistry and Pharmaceutical Sciences, University of Bern, 3012 Bern, Switzerland

[g]Department of Chemistry, Johannes Gutenberg University Mainz, Duesbergweg 10-14, 55128, Mainz, Germany

[h]Department of Chemistry and Department of Physics, The Pennsylvania State University, University Park, Pennsylvania 16802, USA

[#]Current affiliation: Ningbo Institute of Materials Technology & Engineering, Chinese Academy of Sciences, Ningbo, 315201, P.R. China

corresponding author: gabriela.borin-barin@empa.ch

## Abstract

On-surface synthesis enables the fabrication of atomically precise graphene nanoribbons (GNRs) with properties defined by their shape and edge topology. While this bottom-up approach provides unmatched control over electronic and structural characteristics, integrating GNRs into functional electronic devices requires their transfer from noble metal growth surfaces to technologically relevant substrates. However, such transfers often induce structural modifications, potentially degrading or eliminating GNRs' desired functionality - a process that remains poorly understood. In this study, we employ low-temperature scanning tunneling microscopy and spectroscopy (STM/STS) to characterize 9-atom-wide armchair GNRs (9-AGNRs) following polymer-free wet-transfer onto epitaxial graphene (EG) and quasi-freestanding epitaxial graphene (QFEG) substrates. Our results reveal that armchair GNRs maintain their structural integrity post-transfer, while GNRs with extended or modified edge topologies exhibit significant structural changes, including partial disintegration. Additionally, STS measurements reveal differences in the Fermi level alignment between GNRs and the graphene substrates, a key factor in optimizing carrier injection efficiency in electronic transport devices. This study establishes a framework for detecting post-processing structural modifications in GNRs, which are often hidden in optical ensemble measurements. By addressing the challenges of substrate transfer and providing new insights into GNR-substrate interactions, these findings pave the way for the reliable integration of atomically precise GNRs into next-generation nanoelectronic and optoelectronic devices.

## Introduction

The electronic properties of graphene nanoribbons (GNRs) are highly sensitive to their precise chemical structure. For armchair-edged GNRs (AGNRs), changing their width by a single atomic row can dramatically alter their electronic properties, leading to GNRs with bandgaps ranging from quasi-metallic to insulating[1]. Consequently, the integration of GNRs in electronic devices requires their atomically precise synthesis, which can be achieved by selective surface-catalyzed reactions of molecular precursors in highly controlled ultra-high vacuum (UHV) conditions via on-surface synthesis[2]. By carefully designing the precursor monomer, GNRs with armchair[3], zigzag[4], and chiral[5] edge structures, GNR heterojunctions[6], as well as GNRs with topological phases[7,8] have been synthesized with atomic precision.

The most common substrates for atomically precise GNR synthesis via on-surface methods are single crystals of coinage metals, such as Au(111). Once synthesized, GNRs remain adsorbed on these metallic surfaces, enabling their structural characterization using powerful surface analytical tools. In particular, scanning probe microscopy[9] has been instrumental in resolving the exact chemical structures of GNRs, with sensitivity sufficient to resolve atomic defects, such as vacancies or oxidation at specific atomic positions[10].

Among the various types of GNRs, armchair-edged GNRs have been the only class successfully integrated into electronic devices, primarily due to their robustness and chemical stability under ambient conditions.[11] In contrast, GNRs containing zigzag edges or segments are significantly more reactive, as the presence of unpaired electrons in their π orbitals makes them prone to oxidation.[12,13] Scanning tunneling microscopy and Raman spectroscopy studies on metallic growth substrates, conducted under ultra-high-vacuum conditions and controlled oxygen exposure, have shown that zigzag segments in chiral GNRs and at the termini of short AGNRs are the most susceptible to oxidation.[10,14]

Integrating GNRs into functional nanoelectronic devices requires transferring them from the metallic growth substrate to materials compatible with semiconductor processing, such as $Si/SiO_2$ as well as graphene and other 2D materials for contacting and encapsulation. Substrate transfer approaches have predominantly relied on wet-transfer methods, either by using a polymer as a support layer (e.g: PMMA[15], HSQ[16]) or leveraging the Au(111) films on which GNRs are grown on[11]. The latter is the most widely used protocol for integrating GNRs into devices and consists of placing GNR/Au/mica substrates into a hydrochloric solution, which detaches the mica while leaving the GNR/Au film floating on the solution's surface. After successive washing steps with ultra-pure water, the GNR/Au film is transferred onto the target substrate, and the Au film is removed with an iodine-based gold-etching solution. This transfer technique yields higher-quality GNRs compared to polymer-based transfers, where residual PMMA is challenging to remove fully, compromising GNR properties and thus affecting the device's performance[17].

Even for highly optimized protocols, the substrate transfer step remains the main bottleneck in achieving high-performance devices, as it can induce structural modification in the GNRs, potentially degrading or even eliminating the functionalities related to their specific atomic structure. The extent of such modifications has been addressed by Raman spectroscopy investigations where modes related to GNR $sp^2$ lattice, width, and edge geometry can be identified. Previous studies have shown the efficacy of this method in determining GNRs' overall quality and alignment[18–20] on Au substrates and after device integration, as well as determining GNR length[21].

However, since Raman spectroscopy provides ensemble-averaged measurements, the direct identification of specific edge modifications remains elusive. To date, no clear Raman fingerprint of edge-related structural changes has been reported. Moreover, not all GNR moieties produce a Raman signal, and in cases of disintegration, disordered phases may lack a distinguishable optical fingerprint. Yet identifying structural changes between processing steps is essential to optimize the transfer process and retain the desired electronic GNR states that establish transport channels in GNR field-effect transistors (FETs). [22–25] While scanning probe microscopy techniques offer atomic-scale insights into as-grown GNRs, an equivalent level of structural characterization after substrate transfer has not yet been achieved.

In the present study, we transfer exemplary 9-atom-wide AGNRs[26] from their growth substrate onto epitaxial graphene (EG) and quasi-freestanding (hydrogen intercalated) epitaxial graphene (QFEG) on silicon carbide[27], using previously established wet-transfer protocols[11]. EG is an ideal target substrate due to its chemical inertness, temperature stability, UHV compatibility, wafer scale uniformity, and low-temperature conductivity. This allows us to obtain high-resolution STM images and STS spectra of the transferred GNRs on EG and QFEG. Moreover, as graphene is commonly employed as electrical contacts for GNR-based devices, determining the alignment of the electrochemical potential between the GNR and graphene leads is critical for assessing the intrinsic contact resistance. Additionally, we characterize the impact of the transfer process on the structural integrity of GNRs with different edge topologies, including edge-extended GNRs, revealing structural modifications and, in some cases, disintegration. Our STM-based approach thus enables a detailed, atomic-scale evaluation of the impact of the transfer methods on different GNR structures, offering critical insights for integrating atomically precise GNRs into nanoelectronic devices.

## Results and discussions

### Characterization of the 9-AGNR before and after transfer

The integration of atomically precise GNRs into devices presently relies on two main transfer methods: a polymer-free wet chemical transfer[11] and the so-called "bubble transfer"[18]. In the latter, a sacrificial polymer layer (typically PMMA) is used to retain the initial alignment of GNRs during the transfer. Vicinal surfaces, such as Au(788), are typically used to grow aligned GNRs along step edges. The PMMA layer preserves this global alignment, and an electrochemical delamination process releases the PMMA/GNR from the growth crystal, enabling directionally controlled transfer onto the target substrate and high-yield device fabrication[19].

When alignment is not a critical requirement, a more straightforward transfer technique can be employed. GNRs grown on Au(111) thin films on mica substrates are usually transferred using the own Au(111) film as a support layer. After the Au/GNR film is placed on the device substrate, the Au is etched away, leaving GNRs on the substrate (Figure 1a). In this work, we utilize both of these transfer protocols, as described in detail in the methods section.

9-AGNRs are synthesized on 200 nm thick Au(111) films on Mica substrates (Figure 1a) and on a Au(788) single crystal (Figure S1), following established protocols using an automated GNR synthesis system ("GNR Reactor")[11], which ensures reliable inter-sample homogeneity. Different batches of samples prepared in the GNR reactor under identical conditions have consistent quality metrics, such as the average length of the GNRs or the quality of the Raman spectra[11,20]. This ensures a reliable baseline to evaluate post-synthesis processing steps.

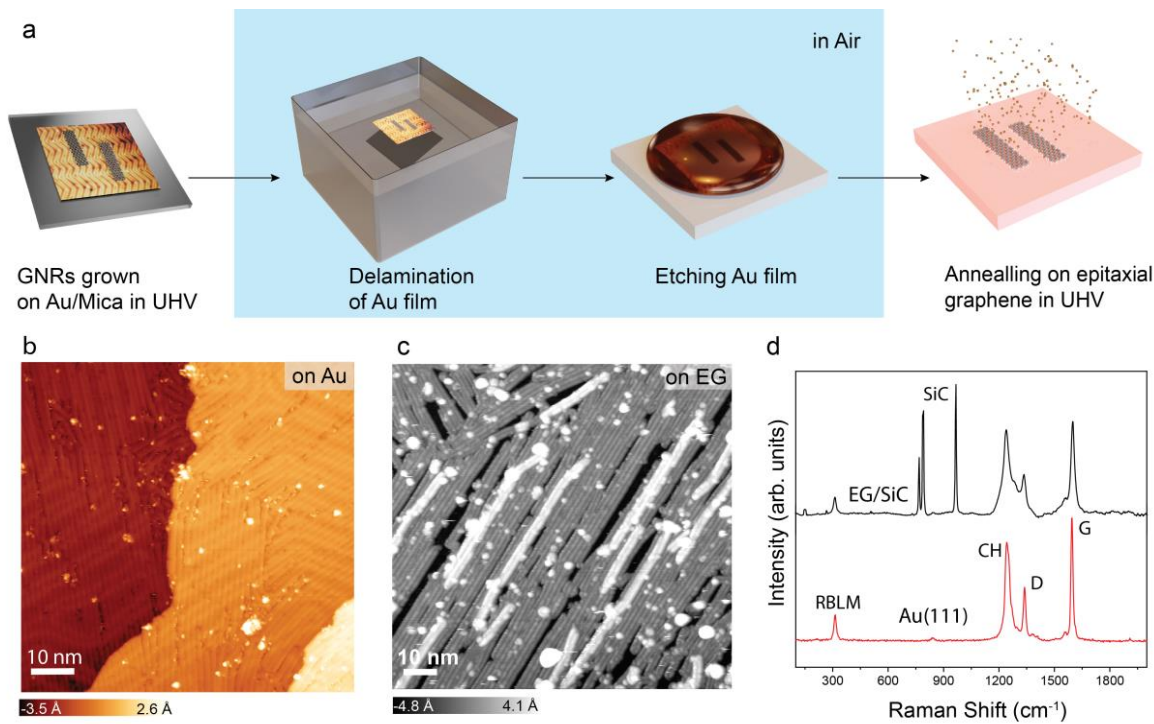

**Figure 1: Transfer of 9-AGNRs onto epitaxial graphene on SiC substrates. a,** Schematic illustration of the protocol used to transfer 9-AGNR grown on Au/Mica surfaces to EG. Once grown, the Au film is delaminated from the mica substrate and is picked up by the EG substrate. The Au film is then etched, leaving the GNRs on the EG substrate. Due to the solution processing steps performed under ambient conditions, the resulting surface is not UHV clean, and a high-temperature annealing (~750 °C) in UHV is required to desorb the surface contaminants. **b**, STM image of the GNRs as synthesized on Au/Mica (-1.5 V, 30 pA). **c**, STM image of the GNRs transferred on EG (1.6 V, 10 pA). **d**, Raman measurements of the GNRs as grown on Au/Mica (red) and after the transfer onto EG and subsequent UHV annealing (black).

After synthesis, GNRs are transferred onto EG by delaminating and etching the gold film for Au/mica substrates and by bubble transfer for the GNRs grown on the Au(788) bulk single crystal (see methods). Previously, ambient characterization of transferred GNRs on $SiO_2$ using atomic force microscopy has shown that the GNRs are transferred as a film, with several GNRs locally aggregating into bundles[11]. However, such samples are non-conductive and typically not clean enough for high-resolution STM inspection. However, sequentially annealing of transferred GNRs on EG substrates to progressively higher temperatures reveals that a clean surface can be recovered after heating to 750 °C in UHV, enabling STM measurements. We presume that these impurities are related to metal salts or polymer residues, both of which would require such high temperatures to decompose and desorb.

We observe excellent transfer uniformity, as evident in pre- and post-transfer STM images, Figure 1b and c (and Figure S1). The measurements indicate that GNRs are mobile during transfer and/or annealing, resulting in some GNRs stacking on top of each other (brighter GNRs in Figure 1c and Figure S1). Statistical analysis reveals a reduction in average GNR length from 26 nm before transfer to 15 nm after transfer (Figure S2). The length distribution remains broad, ranging from 3 nm to 50 nm, with approximately 22% of GNRs exceeding 20 nm. This length reduction, which we tentatively attribute to mechanical/chemical fragmentation during wet transfer and thermal fragmentation during UHV annealing (Figure S2), has important implications for FET applications, where lithography-limited source-drain gaps are typically patterned between 15-20 nm. Only GNRs exceeding this length can effectively bridge the contacts, meaning shorter GNRs and lack of global alignment on Au(111) can contribute to low device yields[28,29]. Raman spectroscopy has been reported to be sensitive to GNR length[21], with the appearance of a mode in the low-frequency region known as the longitudinal compressive mode (LCM). However, the frequency downshift of the LCM is only sensitive to GNRs shorter than 5-7 nm, which is far below the threshold needed to bridge the typical source-drain gaps in FETs. Although we observe shorter GNRs post-transfer, their average length remains above 7 nm, rendering the length decrease undetectable by Raman measurements.

Raman spectroscopy also confirms the relative structural integrity of the transferred GNRs. The radial breathing-like mode (RBLM) at 312 cm$^{-1}$, a fingerprint of the atomically precise width of 9-AGNRs, remains clearly visible before and after transfer. However, the broadening of the CH and D modes, associated with the edge structure and 1D confinement, indicates some degradation, likely due to the non-selective fusion of GNRs during high-temperature UHV annealing. Previous studies have demonstrated that annealing the GNRs on Au(111) surfaces above 450 °C leads to substantial fusion of the GNRs[30], with Raman spectra exhibiting the complete disappearance of the RBLM peak, along with substantial broadening of the G (associated with the GNR sp$^2$ lattice), CH, and D modes. Clearly, such a dramatic change has not occurred in the present sample, despite the significantly higher annealing temperature. This is attributed to the chemical inertness of the EG substrates, which do not catalytically activate the cleavage of the CH bonds. The Raman spectrum after transfer also shows SiC-related modes from the underlying substrate (Figure 1e black spectrum).

High-resolution STM imaging provides direct confirmation that 9-AGNRs are mostly intact after transfer and high-temperature annealing, Figures 2a and b. The choice of the molecular precursor used to obtain the 9-AGNRs has two consequences for the resulting GNR structure. First, it leads to the characteristic armchair-edge (ACE) termini of 9-AGNRs[26]. Second, phenylene rings in the molecular precursor may occasionally be lost during the reaction process, yielding the so-called 'bite-defects' (BD)[26,31]. Both these characteristic features can be seen in the STM image of the transferred GNRs (Figure 2a and b). Additionally, we observe that the ends of some GNRs are fused together post-transfer, an effect we attribute to the high-temperature UHV annealing required to desorb the impurities introduced during the transfer process. Such inter-GNR fusion is not observed in the as-grown sample but remains a relatively minor occurrence after transfer and annealing. Notably, no other apparent edge modification, such as oxidation, was observed.

**Electronic characterization on epitaxial graphene**

The inertness of the EG substrates allowed us to anneal the GNRs to 750°C in UHV to desorb contaminants, and the resulting sample was sufficiently clean to permit low-temperature STS characterization of individual GNRs. As seen in the differential conductance dI/dV spectra shown in Figure 2g, the 9-AGNRs on EG feature well-defined positive (around -1 V) and negative ion resonances (around 0.7 V), attributed to the valence band maximum (VBM) and conduction band minimum (CBM), respectively. Orbital images close to the band onsets, shown in Figure 2c,d, are in excellent agreement with tight-binding simulations (Figure 2e,f), supporting our assignment. The VBM, which is challenging to resolve on metallic surfaces, is most pronounced in the center of the ribbon (light green curve in Figure 2g and Figure S3). STS of 9-AGNRs on QFEG appears qualitatively similar but exhibits a noticeable 210 mV downward shift of the Fermi level relative to EG. This shift originates from the different work functions of EG (4.2 eV) and QFEG (4.7 eV)[32].

The strong correspondence between the experimental orbital maps and gas-phase simulations, combined with the significantly reduced broadening of STS resonances compared to Au, highlights the weak interaction of GNRs with the graphene substrate. Moreover, the 300 mV increase in the measured GNR band gap compared to Au(111)[26] is attributed to the reduced substrate screening. In state-of-the-art transport devices, where GNRs serve as the active transport channel, graphene is typically used as contacts and gate electrode[33]. Hence, the electron and hole addition energies of GNRs measured on graphene substrates closely mimic an actual device configuration (Coulomb energies). Particularly for nano-devices where the separation between the electrodes is only a few nanometers, the level alignment of the channel material is largely determined by the work function of the electrodes when no gate voltage is applied. STS measurements allow us to determine the Fermi level alignment between GNRs and the graphene substrate, providing critical insights into the work function matching of contacts with the GNR transport channel (Figures 2i,j). By analyzing the alignment of the CBM and VBM relative to the graphene Fermi level, we can infer the injection barriers for both charge carrier polarities. This alignment directly impacts the contact resistance, as optimal work function matching minimizes Schottky barriers and enhances charge injection efficiency in device configurations. As shown in figure 2i, GNRs on EG exhibit a more symmetric alignment of VB and CB with respect to the Fermi level compared to GNRs on gold[26] or gold silicide[34]. This symmetry facilitates bipolar transport, allowing for efficient injection of both electrons and holes.

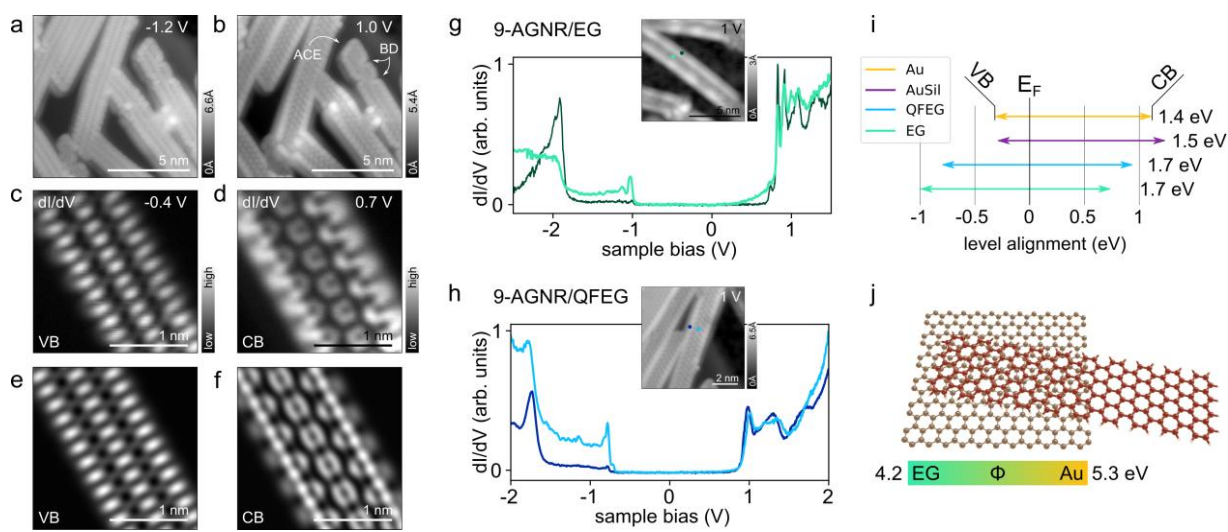

**Figure 2.** Electronic properties of 9-AGNR on EG and QFEG. **a,b** STM topography images of 9-AGNR on QFEG acquired at negative and positive sample bias using a CO-terminated tip. **c,d** CO-tip constant-height STS orbital mapping of frontier VB (c) and CB states (d). **e,f** Tigh-binding simulation of the 9-AGNR VB and CB, assuming a 30%/70% s-wave/p-wave CO-tip character. **g** dI/dV spectra of 9-AGNR on EG at inner (light green) and outer (dark green) ribbon positions (see inset). **h** Differential conductance dI/dV spectra of 9-AGNR on QFEG at inner (light blue) and outer (dark blue) ribbon positions (see inset). **i** Schematic level alignment of 9-AGNR VB and CB on different substrates. The lower work functions of EG (4.2 eV) and QFEG (4.7 eV) with respect to Au (5.3 eV) lead to a more symmetric band alignment around the Fermi level. **j** Model of GNR-graphene heterojunction illustrating the tunability of the contact resistance by the substrate work function.

## Characterization of hybrid and edge-extended GNRs

On-surface synthesis also allows for the incorporation of more complex macrocycles into the GNR structure, yielding, for example, porphyrin-extended zig-zag edged GNRs[35]. The key roadblock preventing the incorporation of such exotic GNRs into devices is their high chemical reactivity. In particular, GNRs with open-shell character are highly sensitive to oxygen exposure, with degradation occurring at pressures as low as $10^{-6}$ mbar[12]. Moreover, Raman signals from reactive GNRs are often weak or entirely quenched due to strong interactions with the metallic substrate. To investigate the feasibility of transferring such chemically reactive GNRs, we applied the same polymer-free transfer method for 9-AGNRs to CoPor-3ZGNRs on EG substrates (see Figure 3a-c). Surprisingly, some of the porphyrin cores survive the harsh chemical (e.g., Au etchant) and thermal treatment. Such information was inaccessible with Raman spectroscopy as these GNRs do not show any Raman signal, both before and after transfer. Possible explanations for that include stronger interactions with the substrate due to the metallic porphyrin cores and band gaps falling outside the resonance window of the available laser excitation wavelengths. While quantitative analysis remains challenging, we observe that several of the cores have pyrolyzed or otherwise degraded (Figure 3a-c). The damage was more pronounced in the case of 7-AGNR-$S$(1,3), known to host topological quantum states[8]. After transfer, no distinct GNR structure could be resolved for these topological GNRs, suggesting significant degradation and loss of atomic precision, Figure 3d-f.

These results highlight the need to develop new transfer protocols tailored to the requirements of chemically sensitive GNRs. Prior to this study, it was well-known that open-shell GNRs exhibit high reactivity, rendering them unsuitable for conventional transfer techniques. Our STM investigations now provide direct evidence that while porphyrin cores can partially survive, topological GNRs suffer severe degradation under current transfer conditions. Promising approaches could include ultra-clean dry transfer techniques, encapsulation in inert layers during transfer, or the use of chemical stabilizers to protect reactive edges and functional groups. These developments are essential for realizing the full potential of chemically sensitive GNRs in device applications.

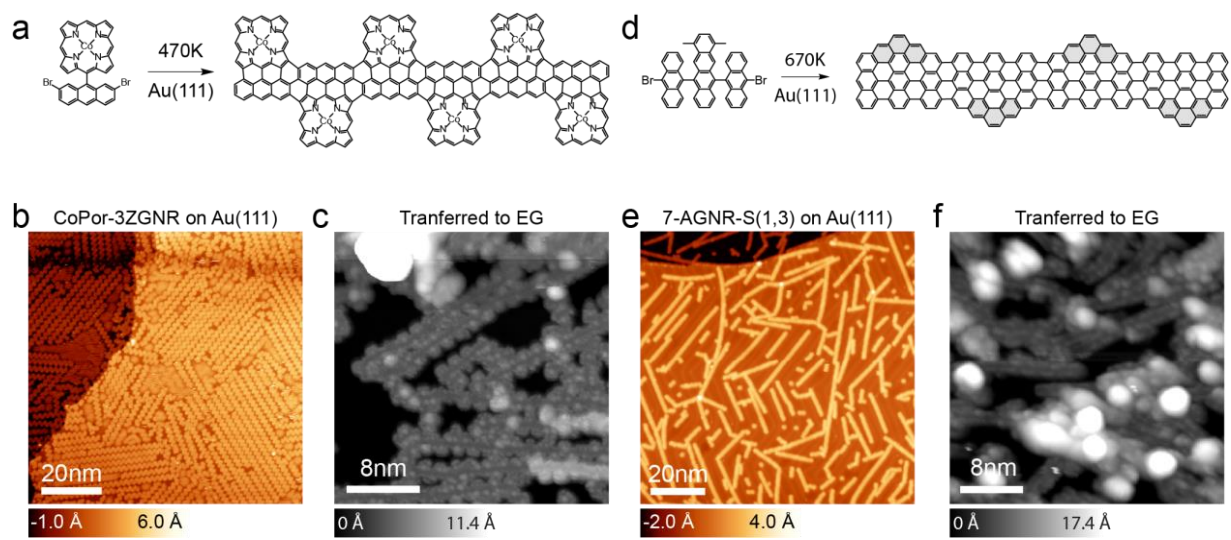

**Figure 3.** STM characterization of hybrid and edge-extended GNRs **a.** STM image of CoPor-3ZGNR as synthesized on Au(111) (-1.0 V, 20 pA) and **b.** after transfer on EG (-2.5 V, 20 pA). **c.** STM image of 7-AGNR-*S*(1,3) as synthesized on Au (111) (-1.5 V, 30 pA) and **d.** after transfer on EG (1.0 V, 50 pA).

## Conclusion

This study provides the first atomic-scale STM evidence confirming the preservation of atomic precision in transferred GNRs, marking a significant milestone in integrating atomically precise nanoribbons into functional devices. While previous studies have relied on ensemble Raman measurements to assess post-transfer GNR integrity, our high-resolution STM imaging directly verifies the retention of atomic structure in chemically stable 9-AGNRs, even after a wet-transfer approach in air. Furthermore, STS and orbital imaging reveal that graphene serves as a weakly interacting substrate, allowing us to assess the intrinsic electronic properties of 9-AGNRs. The significantly increased band gap of 1.7 eV on both EG and QFEG, compared to 1.4 eV on Au, is attributed to reduced screening. Moreover, the symmetric band alignment of the valence and conduction bands with respect to the graphene Fermi level suggests that GNR-based FETs with graphene electrodes are amenable to bipolar charge transport. However, our STM evidence of structural degradation in more reactive GNRs, such as GNRs hosting topological states and porphyrin-core extensions, highlights the urgent need for advanced transfer protocols tailored to chemically sensitive nanoribbons. These results not only solidify the experimental foundation for GNR-based FETs but also open new avenues for exploring GNR-substrate interactions beyond single-crystal metals, expanding the potential of GNRs in quantum transport, spintronics, and optoelectronic applications.

## Methods

*On-surface synthesis of graphene nanoribbons*
Detailed synthesis of 9-AGNRs and 7-AGNR-*S*(1,3) (topological GNR) on Au(111) and Au(788) have been previously published[8,11,19]. A short summary is included here for completeness. 9-AGNRs were synthesized using 3′,6′-di-iodine-1,1′:2′,1″-terphenyl (DITP). DITP was sublimated onto clean Au(111) and Au(788) surfaces in UHV by heating a quartz crucible (sublimation temperature: 70 °C). The 7-AGNR-*S*(1,3) (topological GNR) was synthesized following a similar procedure by depositing the precursor monomer 6,11-bis(10-bromoanthracen-9-yl)-1,4-dimethyltetracene (BADMT) on Au(111) (sublimation temperature: 350 °C). The Co(II)-porphyrin-extended zig-zag GNRs (CoPor-3ZGNRs) were synthesized according to previous reports[35,36]. A short summary is also included here for completeness. The CoPor-ZGNRs were prepared in Au(111) from the precursor Cobalt (II) 5-(2,7-dibromoanthracen-9-yl)porphyrin . The precursor monomer was sublimated onto clean Au(111) in UHV by heating a quartz crucible (sublimation temperature: 380 °C). For all GNR synthesis after precursor sublimation, the gold substrate was annealed to initiate the polymerization (~200°C) and cyclodehydrogenation (300-400°C) reactions. The specific Co-metalated Por-ZGNR precursor is described in the supplementary information (i.e. CoPor-DBA synthesis, figures S4 and S5).

*EG and QFEG growth on silicon carbide*
EG was synthesized on (0001) plane of 6H-SiC (Coherent Corp.) via thermal decomposition of SiC. SiC is firstly annealed at 1400 °C in 10% $H_2$/Ar mixture for 30 min, then heated up to 1800 °C in pure argon and annealed at 700Torr for 20min for the synthesis of monolayer EG. QFEG is prepared via hydrogen intercalation in EG at 950 °C, 600 Torr for 30 min in pure hydrogen.

*Au-Mica transfer*

We transfer the GNRs from the Au thin films on which they were grown to the epitaxial graphene substrates using the well-established delamination followed by etching technique. In brief, the Au on Mica substrates are floated in a small bath of concentrated HCl. HCl serves as an etchant at the interface of Au and Mica. After approximately 20 minutes, the Au thin film is released from the mica substrate, with the latter sinking down into the acid bath and the gold film floating on top of the liquid as a consequence of surface tension. The acid is exchanged with deionized water, and the gold film that remains floating on the liquid is fished out with the target substrate. To ensure a conformal contact between the gold film and the substrate, a few drops of ultra-pure ethanol are added to the fished gold film. This causes the film to stretch out and become planar, and as the ethanol dries, the capillary forces ensure a conformal contact between the thin gold film and the substrate. The drying of the ethanol is aided by heating the sample surface at 100°C for 10 minutes. At this stage, the GNRs are encapsulated between the Au film and the top of the substrate, the present case, epitaxial graphene. The Au film is etched using KI-I gold etchant. The surface of the GNR/EG is thoroughly cleaned by washing in ultra-pure ethanol and acetone, and deionized water. The substrates are then inserted into the UHV system and thoroughly degassed.

*Electrical chemical delamination transfer ("bubble transfer")*

9-AGNRs grown on Au(788) were transferred by electrochemical delamination transfer. In this method, we used poly(methyl methacrylate) (PMMA) as the support layer, spin-coated (4 PMMA layers, 2500 rpm for 90 s) on the 9-AGNR/Au(788) surface. The PMMA/GNR/Au(788) stack was annealed at 80°C (during 10 minutes) prior to delamination. Electrochemical delamination was performed in an aqueous solution of NaOH (1 M) as the electrolyte. A DC voltage of 5 V (current ~0.2 A) was applied between the PMMA/9-AGNR/Au(788) cathode and a glassy carbon electrode used as the anode. $H_2$ bubbles formed at the metallic interface mechanically delaminate the PMMA/GNR layer from the Au(788) surface. The delaminated PMMA/GNR layer was cleaned in ultra-pure water (for 5 minutes) before being transferred to the EG substrate. The PMMA/GNR/EG stack was annealed at 80 °C for 10 minutes + 110 °C for 20 minutes to improve adhesion between the substrate and the PMMA/GNR layer, followed by 15 minutes in acetone bath to dissolve the PMMA. The final GNR/EG was rinsed with ethanol and ultra-pure water. For both transfer procedures the EG substrate was annealed at 450°C for 30 minutes in UHV prior GNR transfer.

*Scanning tunneling microscopy and spectroscopy*

Overview STM images (Figure 1 and S1) were acquired at room temperature in constant current mode using a Scienta Omicron VT-STM. Experimental conditions (sample bias and setpoint current) varied and is reported in the figure captions.

High-resolution images were taken with commercial low-temperature (4.5 K) STMs (CreaTec Fischer & Co. GmbH and Scienta-Omicron GmbH). STM topographic measurements were taken in constant current feedback with the bias voltage applied to the sample. Constant-height dI/dV spectra and maps experimental conditions are reported in the figure caption.

*Raman spectroscopy*

Raman spectroscopy measurements were performed using a WITec confocal Raman microscope (WITec Alpha 300R) with a 785 nm (1.5 eV) laser line and a power of 40 mW. A 50× microscope objective was used to focus the laser beam on the sample and collect the scattered light. The Raman spectra were calibrated using the Si peak at 520.5 cm$^{-1}$. The laser wavelength, power, and integration time were optimized for each substrate to maximize the signal while minimizing sample damage. To prevent sample damage, a Raman mapping approach (10 × 10 μm) was employed.

*Tight-binding calculations*

The tight-binding calculations of the p-wave STM simulations were done on a finite 9-AGNRs (50 anthracene units long) with first nearest-neighbor hopping $t_1$=3 eV. More details on the calculations can be found in Ref. 35[37]

**Acknowledgements**

This work was supported by the Swiss National Science Foundation under grant no. 200020_182015, the European Union Horizon 2020 research and innovation program under grant agreement no. 881603 (GrapheneFlagship Core 3) and the European Innovation Council Pathfinder project no. 101099098 (ATYPIQUAL). The authors also greatly appreciate the financial support from the Werner Siemens Foundation (CarboQuant). R.D. acknowledges funding from the University of Bern. B.S. appreciates funding from the European Research Council (ERC) under the European Union's Horizon 2020 research and innovation program (Grant agreement No. 948243).Y. Gu acknowledges the financial support from the National Natural Science Foundation of China (22405284), the Hundred Talented Program of the Chinese Academy of Science, and the Natural Science Foundation of Zhejiang Province (LZ24B020006) C.D., L.L, J.A.R. acknowledges the National Science Foundations, award number (DMR-2011839 and DMR-1539916). L.P-R. acknowledges the Margarita Salas Postdoctoral Grant supported by Ministerio de Universidades de España and the Resilience Funds Next Generation of the European Union. F.X. thanks the Deutsche Forschungsgemeinschaft (DFG) for a Walter-Benjamin Fellowship (Project No. 452269487) and the SNSF for a Postdoctoral Fellowship (grant No.210093).

# Supplementary Information

# Atomic-scale imaging of graphene nanoribbons on graphene after polymer-free substrate transfer

Amogh Kinikar[a], Feifei Xiang[a], Lucia Palomino Ruiz[a,b], Li-Syuan Lu[c], Chengye Dong[d], Yanwei Gu[e#], Rimah Darawish[a,f], Eve Ammerman[a], Oliver Gröning[a], Klaus Müllen[e,g], Roman Fasel[a,f], Joshua A. Robinson[c,d,h], Pascal Ruffieux[a], Bruno Schuler[a], Gabriela Borin Barin[a*]

[a]nanotech@surfaces laboratory, Empa - Swiss Federal Laboratories for Materials Science and Technology, 8600 Dübendorf, Switzerland.

[b]Departamento de Química Orgánica, Facultad de Ciencias, Unidad de Excelencia de Química Aplicada a Biomedicina y Medioambiente (UEQ), Universidad de Granada, 18071 Granada, Spain

[c]Department of Materials Science and Engineering, The Pennsylvania State University, University Park, Pennsylvania 16082, USA

[d]Two-Dimensional Crystal Consortium, The Pennsylvania State University, University Park, Pennsylvania 16802, USA

[e]Max Planck Institute for Polymer Research, 55128 Mainz, Germany

[f]Department of Chemistry, Biochemistry and Pharmaceutical Sciences, University of Bern, 3012 Bern, Switzerland

[g]Department of Chemistry, Johannes Gutenberg University Mainz, Duesbergweg 10-14, 55128, Mainz, Germany

[h]Department of Chemistry and Department of Physics, The Pennsylvania State University, University Park, Pennsylvania 16802, USA

[#]Current affiliation: Ningbo Institute of Materials Technology & Engineering, Chinese Academy of Sciences, Ningbo, 315201, P.R. China

corresponding author: gabriela.borin-barin@empa.ch

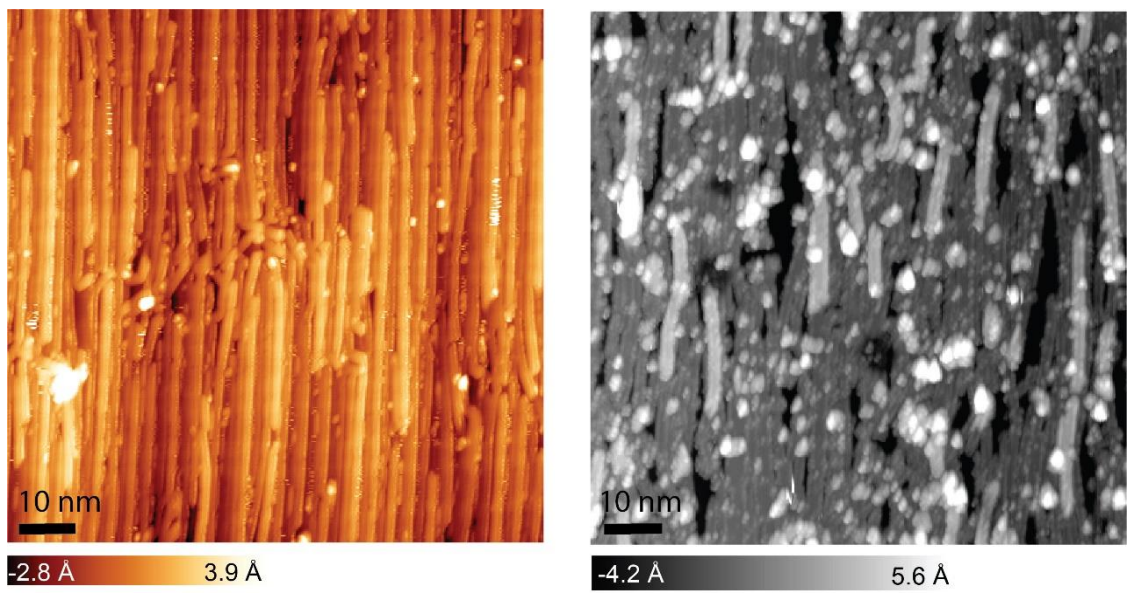

**Figure S1.** (left panel) STM image of aligned 9-AGNRs as synthesized on Au(788) (-1.5 V, 30 pA). (right panel) STM image of aligned 9-AGNRs transferred on EG (-4.6 V, 10 pA).

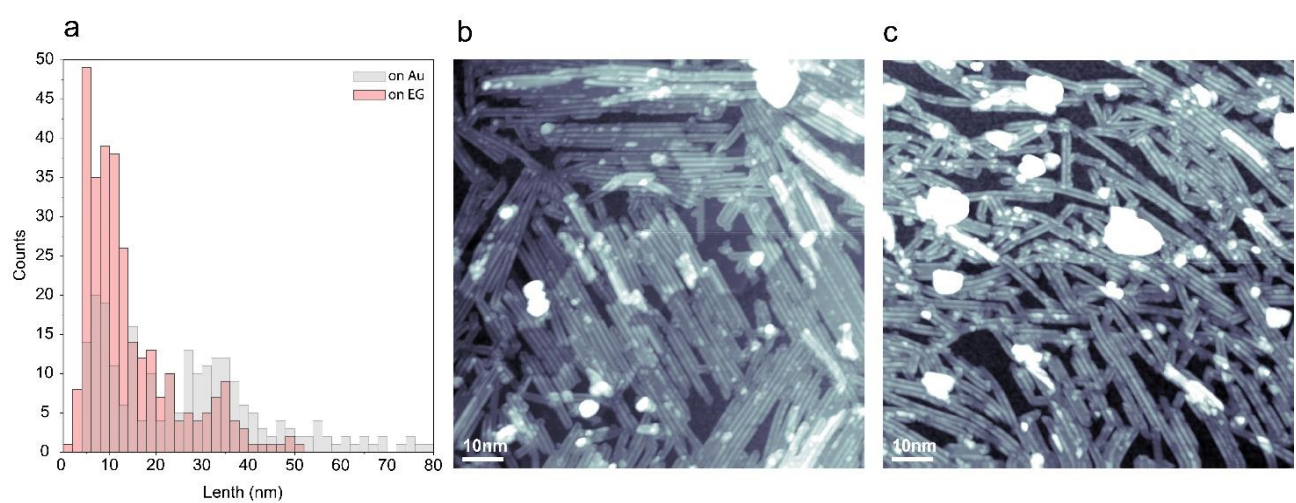

**Figure S2**. a) GNR length distribution before transfer on Au (light grey; figure 1b as a representative image) and after transfer on EG (pink), b-c. Representative 9-AGNR images on EG used for quantifying GNR length. *Images on EG are challenging for statistical analysis, which could lead to underestimating the number of longer ribbons on EG.

The observed reduction in GNR length from an average of 26 nm pre-transfer to 15 nm post-transfer is primarily attributed to mechanical/chemical fragmentation during wet transfer and thermal fragmentation during UHV annealing. Capillary forces and interfacial stresses introduced during delamination, etching, and liquid-based processing likely induce mechanical strain, leading to breakage at intrinsic defect sites. Additionally, the high-temperature UHV annealing step, necessary for impurity desorption, may further contribute to length reduction by inducing thermally driven cleavage at pre-existing weak points in the GNRs. The presence of some fused GNRs suggests that, while annealing promotes impurity removal, it also leads to limited inter-GNR interactions that may influence their final morphology. The absence of oxidation-related modifications in STM and Raman spectra indicates that chemical degradation during gold etching is not a dominant factor. These findings highlight the importance of optimizing transfer and annealing conditions to mitigate structural fragmentation and improve GNR quality for device applications.

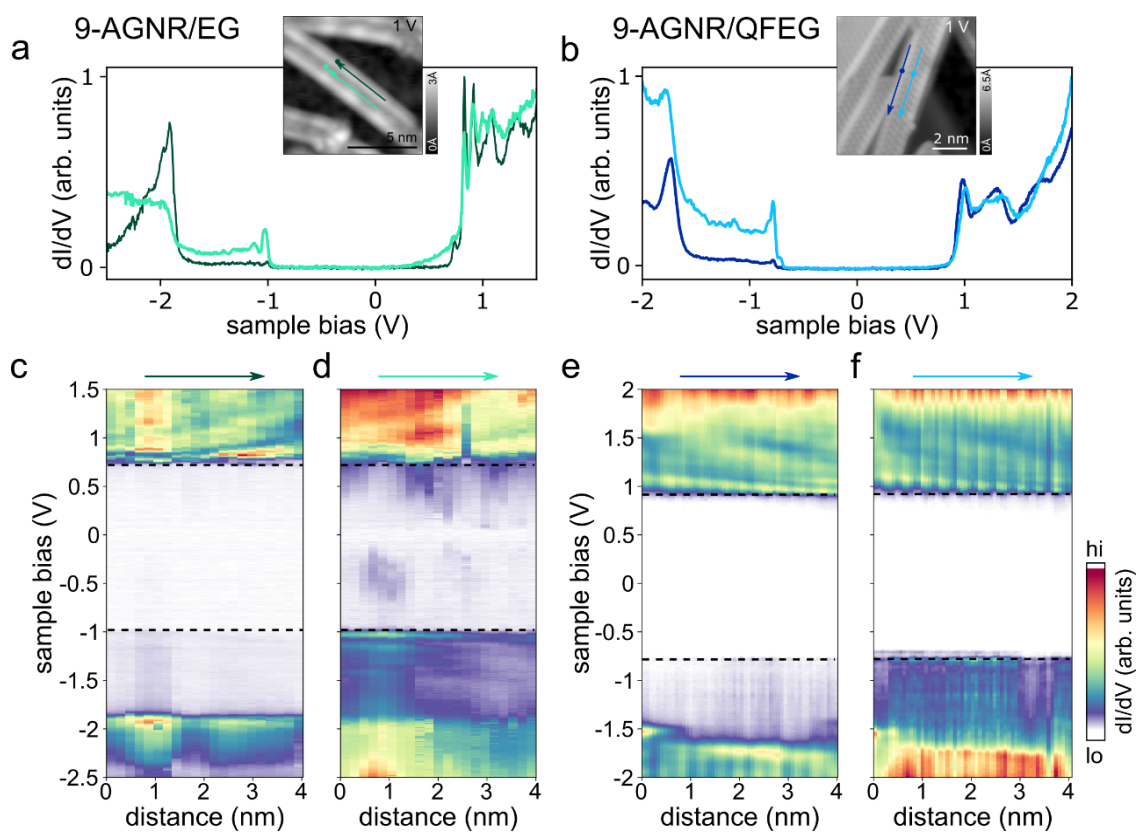

**Figure S3**. STS along 9-AGNR on EG and QFEG. **a** dI/dV spectra of 9-AGNR on EG at inner (light green) and outer (dark green) ribbon positions (see points in inset). **b** dI/dV spectra of 9-AGNR on QFEG at inner (light blue) and outer (dark blue) ribbon positions (see points in inset). **c,d** dI/dV spectra along the outer (c) and inner (d) part of 9-AGNR on EG (see dark and light green arrow in inset in a). **e,f** dI/dV spectra along the outer (e) and inner (f) part of 9-AGNR on QFEG (see dark and light blue arrow in inset in b).

**CoPor-DBA synthesis**

All the reactions were performed in Argon atmosphere supplied into flasks via manifolds. Solvents and reagents were purchased from TCI, Sigma-Aldrich, Acros, Merck, and other commercial suppliers and used without further purification unless otherwise noted. Anhydrous tetrahydrofuran, dichloromethane, and dimethyl sulfoxide were purchased from Sigma-Aldrich and Acros. Column chromatography was conducted with silica gel (grain size 0.063–0.200 mm or 0.04–0.063 mm) and thin-layer chromatography (TLC) was performed on silica gel-coated aluminum sheets with F254 indicator. The high-resolution time-of-flight mass spectrometry (MALDI-TOF) measurements have been performed on a SYNAPT G2 Si high resolution time-of-flight mass spectrometer (Waters Corp., Manchester, UK) with matrix-assisted laser desorption/ionization (MALDI) source. X-ray crystallographic data for precursors **CoPor-DBA** was collected on an IPDS 2T diffractometer using a STOE IPDS 2T diffractometer with a Mo–Kα IμS mirror system radiation. The structure was solved by direct methods SIR-2004 and refined by SHELXL-2014 (full matrix), 450 refined parameters.

Synthetic procedures

Cobalt (II) 5-(2,7-dibromoanthracen-9-yl)porphyrin (**CoPor-DBA**)

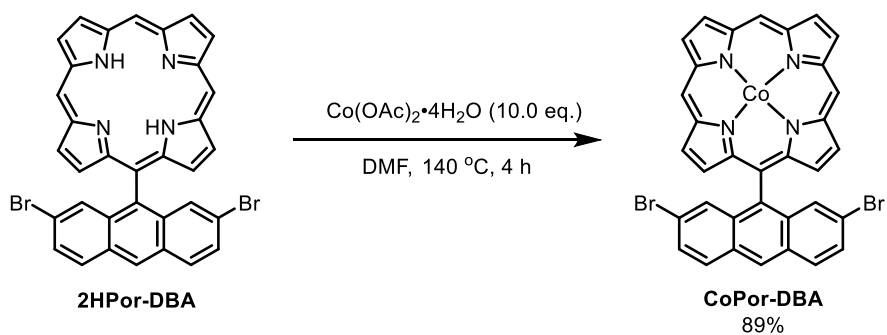

In a 100 mL round bottom flask, cobalt acetate tetrahydrate (124.5 mg, 0.5 mmol, 10.0 equiv.) was added to a solution of compound **2HPor-DBA** (32.0 mg, 0.05 mmol) in *N,N*-Dimethylformamide (30 mL), and the mixture was stirred for 4 hours at 140 ℃. The solvent was removed under vacuum and the residue was purified by silica gel column chromatography (hexane/dichloromethane =7/3) to afford the desired product **CoPor-DBA** (31.2 mg, 89% yield) as pink solid. The further purified compound **CoPor-DBA** for on-surface synthesis was achieved by solvent wash with methanol. Due to its paramagnetic ground state, the proton and carbon nmr were not collected. HRMS (MALDI-TOF, positive) *m/z*: [M]$^+$ calcd for $C_{34}H_{18}Br_2N_4Co$, 698.9230; found: 698.9201. The molecular structure was further confirmed by the single crystal X-ray crystallography.

**Single crystal X-ray diffraction analysis**

The single crystal of precursor **CoPor-DBA** suitable for X-ray analysis was obtained by slow diffusion of methanol into the solution of **CoPor-DBA** in dichloromethane with the exclusion of light. Crystallographic data for compound **CoPor-DBA** is available free of charge from the Cambridge Crystallographic Data Centre under CCDC identifiers 2431713 (www.ccdc.cam.ac.uk/structures/).

**Table S1.** Crystal data and structure refinement for **CoPor-DBA**.

| | |
|---|---|
| Moiety formula | $C_{34}H_{18}Br_2CoN_4$ |
| Formula weight | 701.27 g/mol |
| Temperature | 120(2) K |
| Wavelength | 0.71073Å, MoKα |
| Diffractometer | STOE IPDS 2T |
| Crystal system | Orthorhombic |
| Space group | P bca, (61) |
| Unit cell dimensions | a = 9.6657(7) Å |
| | b = 14.5808(12) Å |
| | c = 37.059(4) Å |
| Volume | 5222.9(8) Å$^3$ |
| Number of reflections | 5787 |
| and range used for lattice parameters | 2.38° <= θ <= 28.17° |
| Z | 8 |
| Density (calculated) | 1.784 Mg/m$^3$ |
| Absorption coefficient | 3.751 mm$^{-1}$ |
| Absorption correction | Integration |
| Max. and min. transmission | 0.8681 and 0.1832 |
| F(000) | 2776 |
| Crystal size, colour and form | 0.040 × 0.070 × 0.760 mm$^3$, brown needle |
| Theta range for data collection | 2.377 to 28.063°. |
| Index ranges | -10<=h<=12, -19<=k<=16, -41<=l<=48 |
| Reflections collected | 14927 |
| Independent reflections | 6235 [R(int) = 0.1771] |
| observed [I>2sigma(I)] | 2133 |
| Completeness to theta = 67.7° | 99.8 % |
| Refinement method | Full-matrix least-squares on $F^2$ |
| Data /restraints / parameters | 6235 / 0 / 370 |
| Goodness-of-fit on $F^2$ | 1.045 |
| Final R indices [I>2sigma(I)] | R1 = 0.1257, wR2 = 0.2688 |
| R indices (all data) | R1 = 0.3033, wR2 = 0.3856 |
| Largest diff. peak and hole | 1.112 and -0.773 e$\text{Å}^{-3}$ |

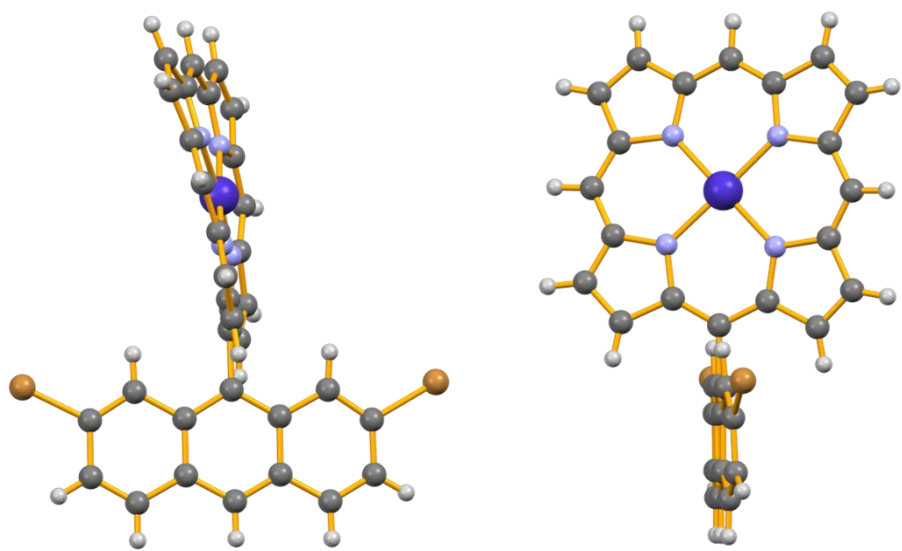

**Figure S4.** X-ray single-crystal analysis of **CoPor-DBA**. Bromine, nitrogen, and cobalt atoms are labelled in brown, purple and blue.

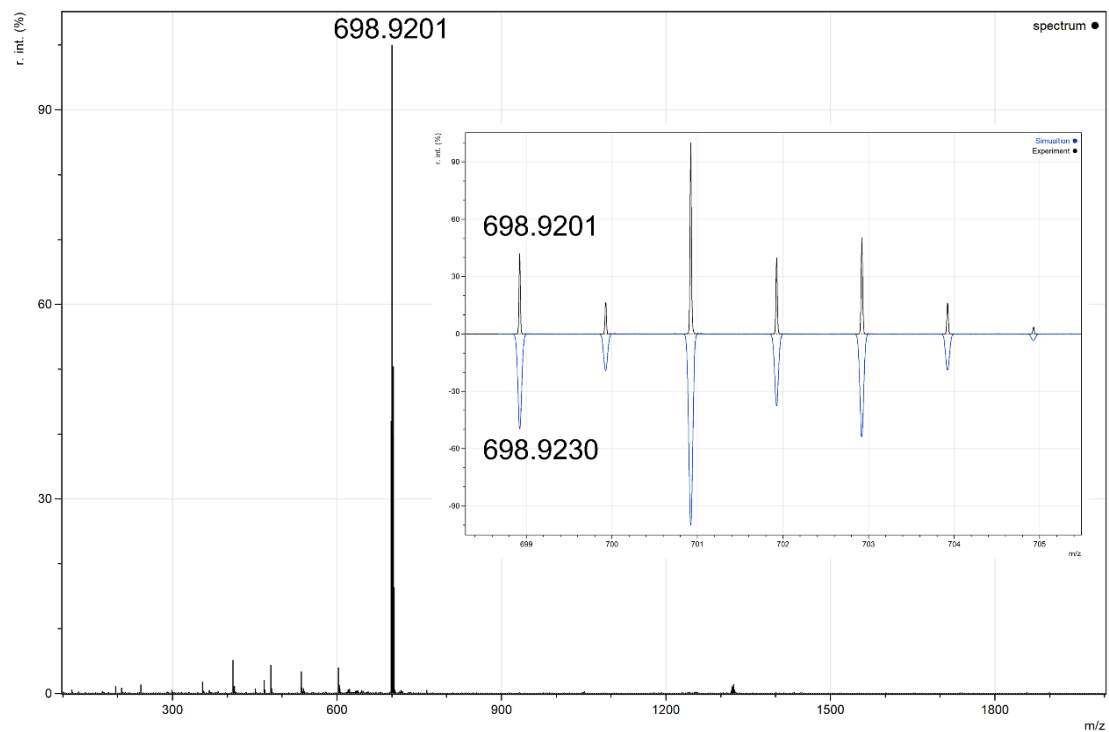

**Figure S5.** High-resolution MALDI-TOF mass spectrum of precursor CoPor-DBA. Inset displays the isotopic distribution (black colour) in comparison to the simulated pattern (blue colour).